\documentstyle[aps,prb,twocolumn,epsfig]{revtex}

\begin{document}
\draft

\twocolumn[\hsize\textwidth\columnwidth\hsize\csname@twocolumnfalse\endcsname
\author{Donavan Hall$^{1}$, E. Palm$^{1}$, T.
Murphy$^{1}$, S. Tozer$^{1}$,
Eliza Miller-Ricci$^{1}$\cite{undergrads}, Lydia
Peabody$^{1}$\cite{undergrads},
Charis Quay Huei Li$^{1}$\cite{undergrads},
U. Alver$^{2}$, R. G. Goodrich$^{2}$, J. L. Sarrao$^{3}$, P.G. Pagliuso$^{3}$, J. M.
Wills$^{4}$, and Z. Fisk$^{1}$}
\address  {$^{1}${\small \ National High Magnetic Field
Laboratory, Florida State University, Tallahassee, FL 32310.}}
\address{ $^{2}${\small Department of Physics and Astronomy,
Louisiana State University, Baton Rouge, LA 70803.}}
\address {$^{3}$ {\small Los Alamos National Laboratory, MST-10, Los
Alamos, NM 87545,}}
\address {$^{4}$ {\small Los Alamos National Laboratory, T-1, Los
Alamos, NM 87545,}}
\title{Electronic Structure of CeRhIn$_{5}$: dHvA and Energy Band
Calculations.}
\date{November 20, 2000}
\maketitle

\begin{abstract}
The de Haas - van Alphen effect and energy band calculations are used
to study angular dependent extremal areas and effective masses of the
Fermi surface of the highly correlated antiferromagnetic material
CeRhIn$_5$.  The agreement between experiment and theory is reasonable
for the areas measured with the field applied along the (100) axis of
the tetragonal structure, but disagree in size for the areas observed
for the field applied along the (001) axis where the antiferromagnetic
spin alignment is occurring.  Detailed comparisons between experiment
and theory are given.
\end{abstract}
\pacs{PACS numbers: 71.18.+y, 71.27.+a}

\vspace{0.5cm}
]

\section{Introduction}

\label{sec:intro}

The compounds CeMIn$_{5}$ (M=Co, Ir, Rh) are a newly reported family
of heavy fermion superconductors 
\cite{Hegger2000,Petrovic2000,Petrovic2000b,Thompson2000}. These materials
crystallize in the tetragonal HoCoGa$_{5}$ structure and are built of
alternating stacks of CeIn$_{3}$ and MIn$_{2}$.  CeCoIn$_{5}$ and
CeIrIn$_{5}$ have superconducting transition temperatures of 2.3 K and
0.4 K, respectively, whereas CeRhIn$_{5} $ orders
antiferromagnetically at 3.8 K at ambient pressure.  Applied pressures
of order 16 kbar can induce an apparently first-order transition from
the magnetically ordered state to a superconducting one with T$_{c}$=
2.1 K. The particular attraction of these materials is that not only
do they represent a substantial increase in the number of known heavy
fermion superconductors but also they appear to be quasi-2D variants
of CeIn$_{3}$, an ambient-pressure antiferromagnet in which
superconductivity can be induced at 25 kbar and 100
mK.\cite{GGL-Nature} If one can demonstrate that the reduced
dimensionality is responsible for the factor of ten increase in
superconducting T$_{c}$ as well as the anomalous evolution from
antiferromagnet to superconductor with pressure, their impact will far
exceed that of just another family of heavy fermion superconductors. 
However, much work remains to establish this hypothesis.

Because it displays both `high-temperature' superconductivity as well
as unconventional magnetic behavior, CeRhIn$_{5}$ seems an ideal
candidate for electronic and magnetic spectroscopic probes to
establish the degree of reduced dimensionality and its impact on
fluctuation spectra.  Some progress has been made in establishing the
magnetic structure of CeRhIn$_{5}$\cite{Curro2000,Bao2000}.  Nuclear
quadrupole resonance and neutron diffraction measurements have shown 
that
the magnetically ordered state of CeRhIn$_{5}$ is a spiral spin
structure in which the spins in a given CeIn$_{3}$ plane are ordered
antiferromagnetically, and the direction of the ordered moments in the
plane spirals along the c-axis, rotating 107 degrees per layer. 
Although complete inelastic neutron scattering measurements of the
dynamical susceptibility remain in progress, already NQR has
established that the temperature evolution of the sub-lattice
magnetization below T$_{N}$ is much more rapid than would be expected
from mean field theory and is a sign of reduced dimensionality.

To date, studies of the electronic structure of CeRhIn$_{5}$ have been
far less complete.  Cornelius {\em et al.} \cite{Cornelius} have
reported limited de Haas-van Alphen (dHvA) measurements along the
principal crystallographic axes using pulsed fields.  These measurements reveal an
anomalous temperature-dependence of the dHvA amplitude as well as
large directional anisotropies in the effective masses of carriers
that have been observed.

Here, we report a more comprehensive dHvA
study throughout the entire Brillouin zone of CeRhIn$_{5}$ and compare
our results directly with local-density-approximation band structure
calculations.  Recently, a similar study of CeIrIn$_{5}$ has been
reported.\cite{Haga2000}  Taken together, our results and those of
Reference \onlinecite{Haga2000} allow direct evaluation of the changes in
electronic properties of CeMIn$_{5}$ as a function of transition metal
ion M.

In what follows, we report the details of our dHvA measurements and
band structure calculations, discuss the extent to which they are
self-consistent, and compare and contrast the electronic structures of
CeRhIn$_{5}$ and CeIrIn$_{5}$.

\section{Measurements}
\label{sec:meas}

All of the measurements reported here were made at the National High
Magnetic Field Laboratory, Tallahassee, FL using cantilever magnetometry at
temperatures between 20 and 500 mK in applied fields ranging from 5 to 18 T.
Complete field rotations in the [100] and [001] planes of the tetragonal
structure are reported. The sample was
grown from an In flux and etched in a 25\% HCl in H$_{2}$O solution down to a
small plate that was mounted on the cantilever with Apiezon N vacuum grease
or GE Varnish. This technique is most sensitive to low frequency
oscillations not observed typically with the magnetic field modulation 
technique.

The dHvA effect, a measurement of the oscillatory part of the magnetization,
is a method for determining Fermi surface properties. The oscillatory
magnetization $M$ is given by the Lifshitz-Kosevitch (LK) equation (see Ref.
\onlinecite{Shoenberg} for the mathematical details):
\begin{eqnarray}
M &=&-2.602\times 10^{-6}\left( \frac{2\pi }{HA^{\prime \prime }}\right)
^{1/2}  \nonumber \\
&&\times \frac{GFT\exp (-\alpha px/H)}{p^{3/2}\sinh (-\alpha pT/H)}\sin
\left[ \left( \frac{2\pi pF}{H}\right) -\frac{1}{2}\pm \frac{\pi }{4}\right]
,  \label{eq:LK}
\end{eqnarray}
where $\alpha $ = 1.47(m/m$_{0}$)$\times $10$^{5}$ G/K, $A^{\prime \prime }$
is the second derivative of the area of the Fermi surface (FS) cross-section
that is perpendicular to the applied field, G is the spin reduction factor, $
p$ is the harmonic number, and $x$ is the Dingle temperature. The
frequencies of the dHvA oscillations are proportional to extremal areas of
the FS, and Fermi-liquid theory works well for heavy-fermion materials as
has been shown in previous studies.\cite{Wasserman1996}

Since the magnetization of the sample was measured with a torque cantilever,
it is necessary to understand how torque is related to the LK equation. dHvA
oscillations in the torque arise from anisotropy in the Fermi surface, such
that
\begin{equation}
\tau =\frac{-1}{F}\frac{dF}{d\theta }MHV  \label{eq:LKtorque}
\end{equation}
where $F$ is the dHvA frequency, $\theta $ is the angle of the applied
field, $M$ is the LK expression above, and $V$ is the volume of the sample.
This means that a roughly spherical FS will have a smaller torque signal
than a highly elliptical FS.

\begin{center}
\begin{figure}
\epsfig{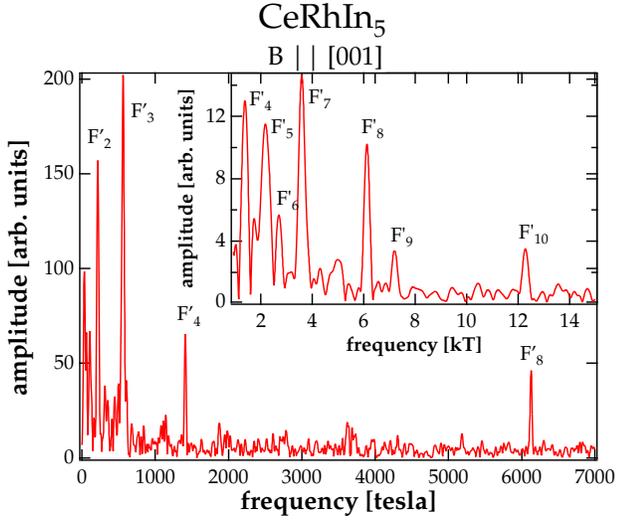}
\caption{FFT of dHvA
data taken with the applied field parallel to the [001].}
\label{fftraw001}
\end{figure}
\end{center}

Two versions of a Fourier transform algorithm were used in the
analysis: preliminary analysis was done with a fast Fourier transform
(FFT) and in some cases a discrete Fourier transform (DFT) was used to
increase the frequency resolution.  Fourier spectra of the
oscillations obtained for the field parallel to [100] and [001] are
shown in Figure \ref{fftraw100} and Figure \ref{fftraw001}.  As can be
seen there are several fundamental frequencies and several
combinations (especially with the F$_{7}$; e.g. F$_{7}$ - F$_{6}$ =
1302 T).  Similar combination frequencies were found in all field
directions; their origins will be discussed below.

\begin{center}
\begin{figure}
\epsfig{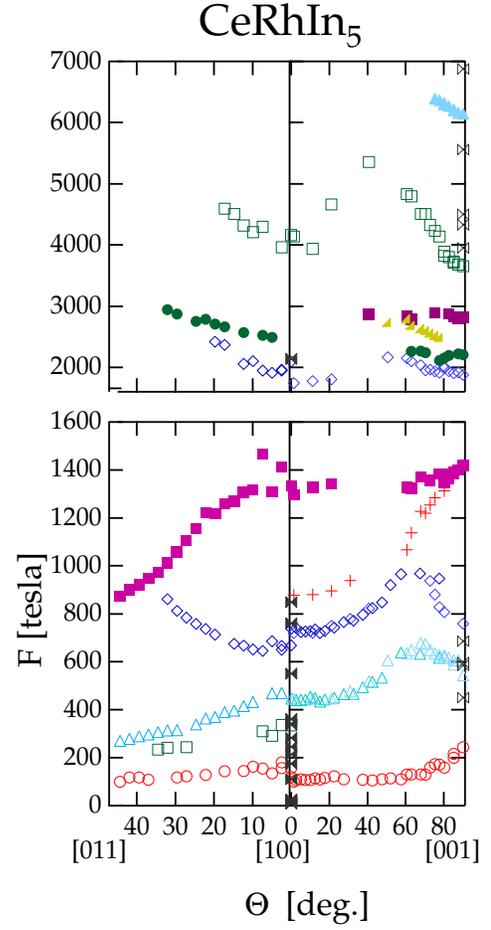}
\caption{Frequency as a function of angle.  Calculated frequencies 
for [001] and [100] are indicated with bowties.}
\label{rotcomp}
\end{figure}
\end{center}

The fundamental frequencies are plotted as a function of angle in
Figure \ref{rotcomp}.  As can be seen the measured frequencies span
the range from about 100 T over wide angular ranges to 5 kT for the
field applied near [001].  The one defect of the cantilever
technique was alluded to in the previous paragraph -- small
$\frac{dF}{d\theta }$ produces a small signal amplitude.  The measured dHvA
oscillations were small for [001] and [100] orientations, and the
signal amplitude was maximized at approximately a forty-five degree
rotation from each of these principal axes. 
One might propose that field modulation measurements would be more
appropriate in this case, but given the low (less than 1 kT)
frequencies measured here, the required modulation fields are
technically impractical in the high field ranges used.

\section{Discussion}

\label{sec:discuss}



The total energy and band structure of CeRhIn$_{5}$ were calculated
with a full-potential electronic structure method that uses linear
muffin-tin orbitals as bases.\cite{wills_es} Calculations were
performed to obtain theoretical structural parameters, one-electron
bands and spectral densities, and the structure of the Fermi surface. 
All calculations used the GGA \cite{PBE} to treat exchange and
correlation and included the spin-orbit interaction in the variational
basis.  In the range of parameters searched, no magnetic instabilities
were found.  The calculations reported here are for paramagnetic
CeRhIn$_{5}$ in the simple tetragonal structure (P4/mmm).

Structural parameters were obtained by simultaneous variation of
volume, $ c/a $, and the single internal structural parameter to
minimize the total energy.  Substantially converged results were
obtained with a set of 45 irreducible (486 total) Brillouin zone
points.  The calculated volume, $c/a$ ratio, and internal parameter
are given in Table \ref{band-table}.  Conventional electronic
structure theory (with itinerant Ce 4$f$ electrons) applied to a
simple paramagnetic cell seems to give a satisfactory description of
the structural properties of CeRhIn$_5$.  Also given in Table
\ref{band-table} are the bulk modulus calculated at both the
experimental and theoretical volumes and the Density of States (DOS)
at the Fermi energy ({\rm E$_{F}$}).  calculated at the experimental
volume.

\begin{center}
\begin{figure}
\epsfig{file=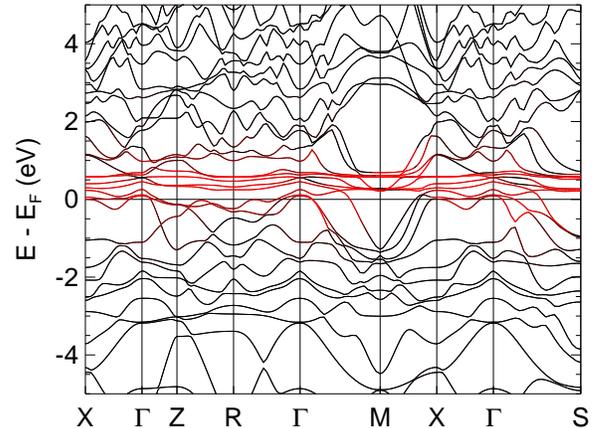, height=.35\textwidth}
\caption{The GGA band
structure of paramagnetic CeRhIn$_5$ calculated using experimental
structural parameters.}
\label{band-structure}
\end{figure}
\end{center}

\begin{center}
\begin{figure}
\epsfig{file=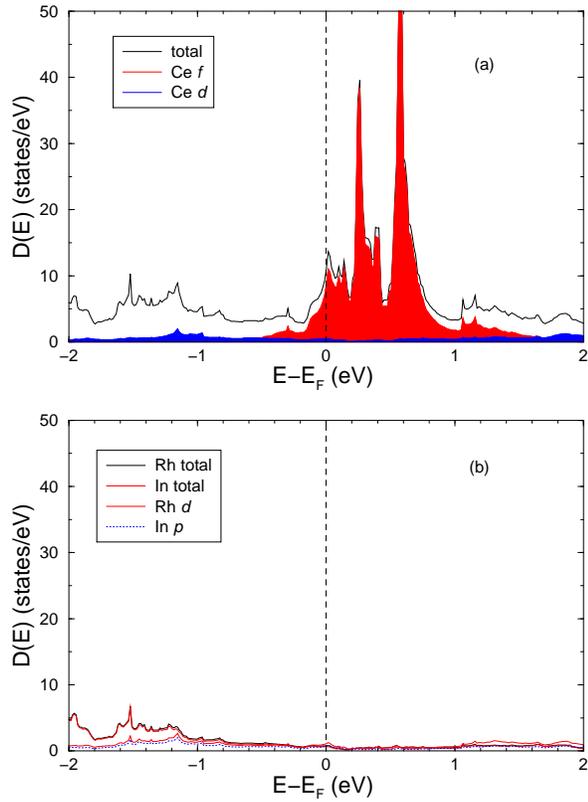, height=.6\textwidth,clip=}
\caption{The GGA density of states of paramagnetic CeRhIn$_5$ calculated using
experimental structural parameters.
Ce 4$f$ and 5$d$ projections are shown in (a), and Rh and In
projections in (b).
}
\label{dos}
\end{figure}
\end{center}

The band structure, calculated at the experimental volume, along high
symmetry lines in the Brillouin zone is shown in Figure
\ref{band-structure}.  Bands of predominantly Ce 4$f$ character
dominate the band structure at and just above {\rm E$_{F}$}.  The set
of bands along $X \to \Gamma \to Z \to R$ are indicative of the
nesting found in the Fermi surface discussed below.  The electron DOS,
shown in Figure \ref{dos}, shows the predominance of Ce 4$f$ character
at the Fermi energy.  The DOS at {\rm E$_{F}$}, while fairly large,
does not indicate a dramatic structural instability.

\begin{center}
\begin{figure}
\epsfig{file=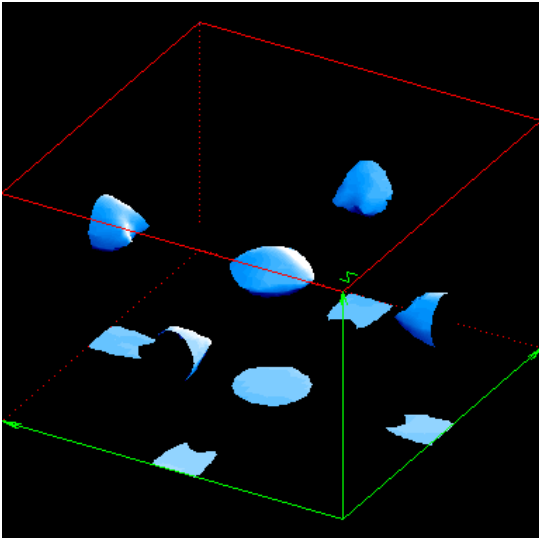, height=.4\textwidth,clip=}
\caption{
The first sheet of the paramagnetic CeRhIn$_5$ Fermi surface.
The electron surface is illuminated; these are hole surfaces.
}
\label{sheet1}
\end{figure}
\end{center}

\begin{center}
\begin{figure}
\epsfig{file=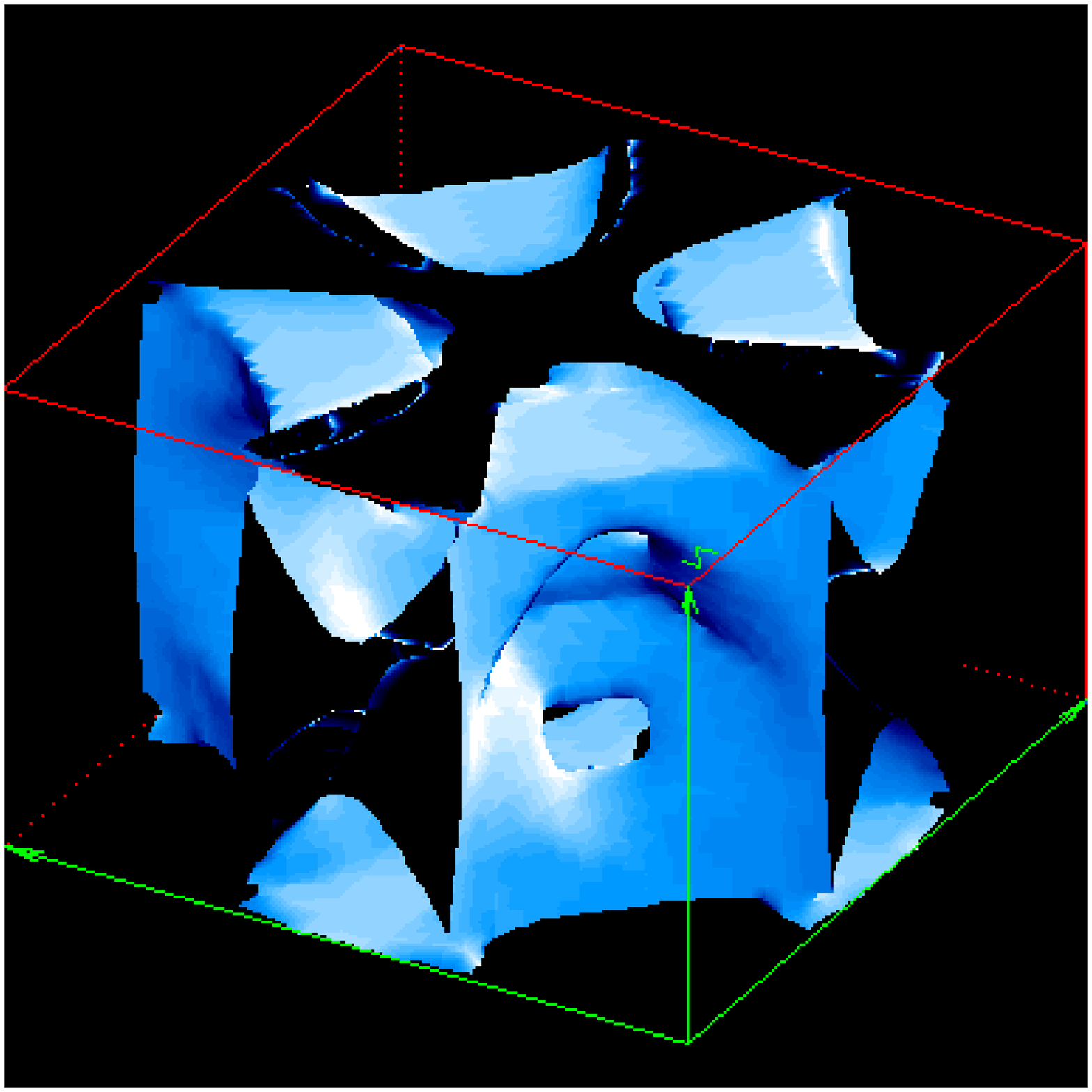, height=.4\textwidth,clip=}
\caption{
The second sheet of the paramagnetic CeRhIn$_5$ Fermi surface.
The electron surface is illuminated.
}
\label{sheet2}
\end{figure}
\end{center}

\begin{center}
\begin{figure}
\epsfig{file=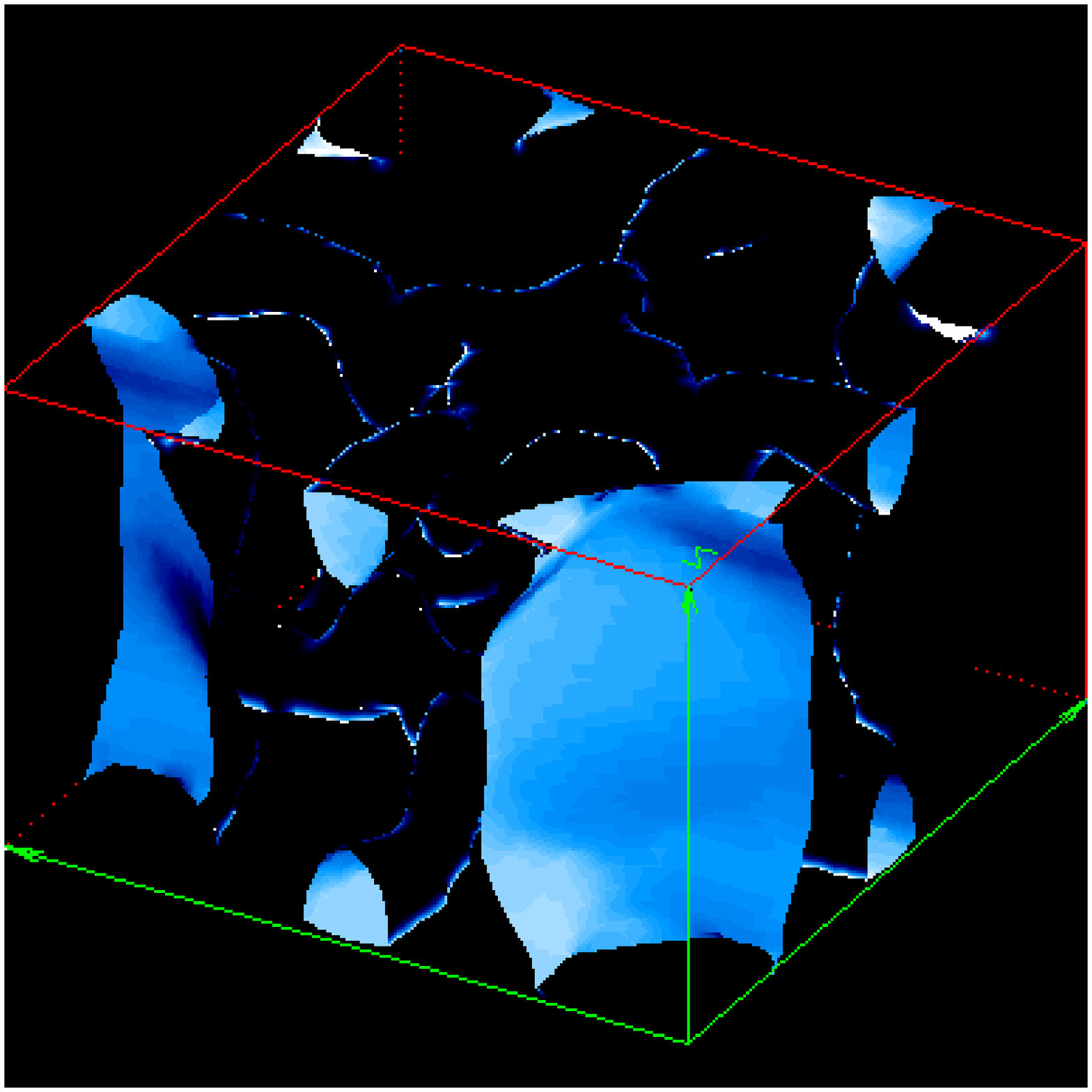, height=.4\textwidth,clip=}
\caption{The third sheet of the paramagnetic CeRhIn$_5$ Fermi surface.
The electron surface is illuminated.}
\label{sheet3}
\end{figure}
\end{center}

Fermi surfaces were calculated using a fine mesh (4800 irreducible
points, corresponding to a 48 $\times$ 48 $\times$ 32 gridding of the
Brillouin zone) with a potential converged with 270 irreducible
Brillouin zone points.  Three doubly degenerate bands, which we label
band 90, band 92, and band 94, cross the Fermi energy.  The Fermi surface
sheets formed by these three bands are shown in Figures \ref{sheet1},
\ref{sheet2}, \ref{sheet3}.  In the
figures, $\Gamma$, the center of the conventional Brillouin zone, is
at the origin, {\it i.e.} at the corners of the cubes.  The lowest
energy sheet, band 90, shown in Figure \ref{sheet1}, consists of hole
surfaces centered around $\Gamma$ and $X$, giving relatively low
(${\buildrel < \over \sim}$ 600 T) dHvA frequencies.  The next sheet,
shown in Figure \ref{sheet2}, has both hole and electron surfaces
stretched along the tetragonal axis.  The last sheet, shown in Figure
\ref{sheet3}, is predominantly an electron surface, although there is
a large area ([001]) hole surface on that sheet.  This sheet
consists of cylinders surrounding [001] axis and a lattice-structure
surface.  The dHvA frequencies corresponding to extremal 
areas are given
in Tables \ref{bandfreqs100} and \ref{bandfreqs001}.

We measured the temperature dependence of the dHvA amplitudes in the
primary crystallographic directions.  We did not observe the anomalous
temperature-dependence of the dHvA amplitudes noted in Ref. 
\onlinecite{Cornelius}, where for B $\parallel$ [001] the amplitudes of the
oscillations increase with increasing temperatures between 0.4 and 1.2
K then decrease.  Our measurements, carried out between 25 and 500
mK, show decreasing amplitudes with increasing temperature in all cases.

The measurements reported here agree substantially with Cornelius \emph{et
al.}\cite{Cornelius} after a 5\% correction to their
data\cite{CorneliusPC}; all but one (4708 T) of the frequencies
reported in Ref. \onlinecite{Cornelius} also are observed here.  One major
difference is the measurements along [001] where we see many more
frequencies.  Our measured
frequencies in this direction run low as compared to calculations, but
we share two frequencies measured by Cornelius \emph{et al.}: our
F'$_{7}$ and F'$_{8}$ (see Table
\ref{bandfreqs001}\cite{CorneliusPC}).  Also, the masses we measure
for these orbits are in excellent agreement (see Table
\ref{cmass001}).  Cornelius \emph{et al.} report a high mass orbit at
4686 T which we do not see.  Another major difference is our measured
masses for [100] are higher when compared with those of Cornelius
\emph{et al.} (see Table \ref{cmass100}), a fact consistent with our 
non-observance of the anomalous temperature dependence.

\section{Comparison of Experiment with Theory}

\label{sec:compare}

\begin{center}
\begin{figure}
\epsfig{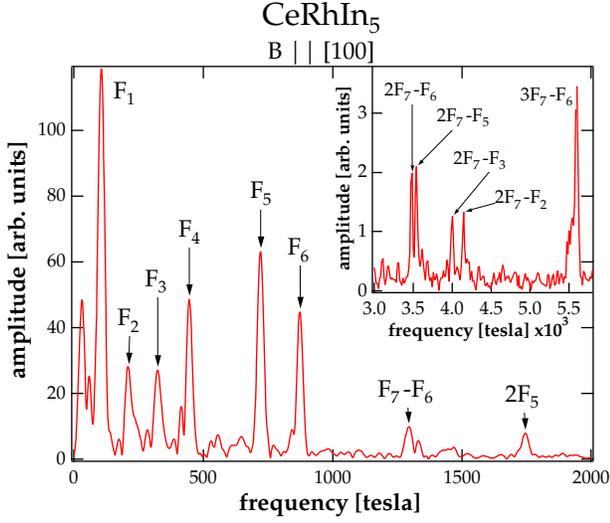}
\caption{DFT of dHvA data taken with   the applied 
field parallel to the [100].  The inset
shows some of the high frequencies that are combinations of the
fundamentals.}
\label{fftraw100}
\end{figure}
\end{center}

We start the comparison between the calculated frequencies and the
measured frequencies with those calculated and observed for the field
along the [100] direction.  In this case there are many more measured
frequencies than the calculated FS gives.  We attribute this situation
to strong magnetic interactions, between electrons on different parts
of the FS, or to torque interactions due to the use of the cantilever. 
Either of these effects can give rise to a series of frequencies that
are combinations of two fundamental frequencies.\cite{Shoenberg} The
measured fundamental frequencies in Table \ref{bandfreqs100} are
indicated as F$_{1}$ through F$_{7}$, and some of the combinations
that are differences between the fundamentals or harmonics of a
fundamental and another fundamental frequency are shown in Figure
\ref{fftraw100}.  It should be noted that the same fundamental
frequencies are subtracted from both the fundamental 2176 T and the
first two harmonics of 2176 T (F$_{7}$) to give the observed
combination frequencies (see the inset in Figure \ref{fftraw100}).  In
one case, 1744 T is the first harmonic of the fundamental 874 T
(F$_{6}$) frequency.

Five frequencies ranging from 9 to 26 T obtained in the band calculations
are too low to be observed experimentally. That is, in the
field range where oscillations are observed (from about 7 to 18
T) there are too few oscillations for these frequencies to be 
measurable. In addition there are four electron orbits grouped into two pairs,
245 T and 215 T plus 175 T and 110 T, for which we observe only two
frequencies, 212 T and 106 T. Again, the Fourier transform may not have sufficient
resolution due to an insufficient number of oscillations to resolve all of
these frequencies.

The remaining five measured frequencies are in the same range as the
predicted frequencies from the band calculation. Differences in detail
occur, but it appears that for the field in this direction the measurements
give strong support to the predicted FS.

The agreement between the calculated and measured frequencies is not as good
for the field in the [001] direction. There are three calculated high
frequencies, 11.268 kT, 12.126 kT, and 12.326 kT, whereas only one frequency
in this range is observed, 12.280 kT. This high frequency only is observed
in the lowest temperature data, 20 mK, and indicates a large effective mass
for this frequency. The fact that only one rather than three frequencies are
observed may indicate that the electron surface is much more cylindrical
than is obtained from band theory. For the remainder of the frequencies,
only the total number and order of the observed frequencies agree with the
calculations.

In the band calculation the fact that the sample is antiferromagnetic over
all of the measurement field ranges has not been taken into account. In the
model that has been taken for the band calculations, the Ce 4{\it f}
electron is completely itinerant and there are no localized moments to cause
the magnetism. Therefore, it is most likely that spin density waves exist in
the electronic system and drastically alter the FS. This is the accepted
situation in Cr,\cite{Fawcett} and may be the case here. A much more involved theory is
required to predict the FS under these circumstances. Exchange interactions
could cause the FS to be split into separate spin up and spin down sheets.
In order for this to be determined from dHvA measurements several harmonics
of these frequencies would have to be measured.\cite{Alem} We did not
observe the harmonics in these measurements, and need to make measurements
to higher fields at mK temperatures to do so.

\section{Comparison of CeRhIn$_{5}$ with CeIrIn$_{5}$}\label{sec:rhir}

The structure of the Fermi surface of CeRhIn$_{5}$ is essentially the same as that
calculated for CeIrIn$_5$ by Haga {\it et al.} \cite{Haga2000}
except that the frequency of the hole orbit perpendicular to [001]
arising from (what we label) band 94 near $\Gamma$ is $\sim$ 11000 T,
rather than $\sim$ 15000 T reported in Reference \onlinecite{Haga2000}.  Our
own calculations for CeIrIn$_5$ find a frequency close to that
calculated for CeRhIn$_5$, $\sim$ 11600 T\cite{wills-unpub}.  The high frequency
reported in Reference \onlinecite{Haga2000} was not seen experimentally. 
The difference in the two calculations may be in our inclusion of the
spin orbit interaction or in the fineness of the grid used to
calculate the Fermi surface.

The dHvA measurements on CeIrIn$_{5}$ \cite{Haga2000} find eight
branches for rotations in the [100] and [110] planes of the tetragonal
structure.  Most of these branches are associated with large quasi-2D
undulating cylinders that show the expected 1/cos($\theta$) dependence 
of the measured frequencies
with $\theta$ being the angle at which the field is applied away from
the [100] axis.  We see a similar behavior in the high frequency
orbits for CeRhIn$_{5}$.  Band structure calculations predict, in Ref. 
\onlinecite{Haga2000} and in Ref.  \onlinecite{wills-unpub}, in
addition, that several small pieces of FS should exist in both
CeRhIn$_{5}$ and CeIrIn$_{5}$.  These frequencies were not observed in
Ref.  \onlinecite{Haga2000} for CeIrIn$_5$ but are observed here in
CeRhIn$_{5}$.  These differences likely are a result of the measurement
techniques employed and not due to differences in the actual Fermi
surfaces of the respective materials.  Overall, the result of these
angular dependent dHvA measurements on CeRhIn$_{5}$ along with the band
calculations show that the Fermi surface has both small pockets of 3D
character and large undulating cylinders that should give rise to a 2D
character in other properties.

From the temperature dependence of the signal amplitudes we have
calculated the effective masses of the observed frequencies in cases
where signals were observable over the entire temperature range of
measurement.  There are two cases where no masses are reported here:
(1) for F'$_{1}$ - F'$_{3}$ in Table \ref{cmass001} the masses are
sufficiently light that we observe no amplitude change within
measurement uncertainty up to 0.5 K, and (2) for F$_{7}$ in Table
\ref{bandfreqs100} and F'$_{5}$, F'$_{6}$, F'$_{9}$ and F'$_{10}$ in
Table \ref{bandfreqs001}, the signals are observed only at the lowest
temperatures.  The masses are obtained from fits to the Dingle
reduction factor in Eq. \ref{eq:LK}:
\begin{equation}
	R_{T} = \pi \lambda/sinh(\pi \lambda),	
	\label{eq:Dingle}
\end{equation}
where $\lambda = 2\pi pkT/\beta B$, $\beta = eh/(m/m_{0})c$, and $m/m_{0}
= m*$ is the effective mass. 
All of the effective masses we obtain in this manner are greater than
1, and compare favorably to the values found in Ref. \onlinecite{Cornelius}
from the high temperature dHvA data.  We note that none of the
measured masses are sufficiently large to account for the measured
value of gamma of 400 mJ/mole-K from high temperature specific heat
measurements\cite{Cornelius}.  On the other hand there is a loss of
entropy due to magnetic ordering, and the inferred value of gamma
below T$_{N}$ is 56 mJ/mole-K, much more in line with the masses reported
here.  The effective masses measured on CeIrIn$_{5}$\cite{Haga2000} are
larger than we observe in CeRhIn$_{5}$ consistent with the
larger value of gamma measured in CeIrIn$_{5}$\cite{Petrovic2000} (in 
which no magnetic order is observed).

\section{Conclusion}\label{sec:conclusion}

We have determined the electronic structure of CeRhIn$_{5}$.  This
determination includes both the results of dHvA measurements and
energy band calculations.  The major findings are:

\begin{itemize}
\item The Fermi surface consists of structures arising from two hole
bands and one electron band.  The band calculations assume a
completely itinerant 4{\it f} electron from the Ce atoms and the
measurements confirm this assumption.  One of the hole surfaces and
the electron surface are open along the c - axis and could give rise
to some rather two-dimensional character to the electronic properties.

\item  While the energy band calculations are for paramagnetic CeRhIn$_{5}$
and the measurements are on anti-ferromagnetic CeRhIn$_{5}$, the agreement
between experiment and theory is good for electron areas perpendicular to
the a-b plane, but deviate in size for orbits in this plane. This
disagreement is thought to be due to spin density waves in the c axis
direction that give rise to the anti-ferromagnetism in the a - b plane.

\item The electronic effective masses that are measured all are
greater than the free electron mass and are anisotropic, varying from
less than m$_{0}$ to 6m$_{0}$ depending on direction.
\end{itemize}

Overall the Fermi surface of anti-ferromagnetic CeRhIn$_{5}$ is
similar to that found previously \cite{Haga2000} in superconducting
CeIrIn$_{5}$.  Thus, while their ambient pressure ground states
are completely different, both of these materials have large electron - electron
interactions and rather similar electronic
structures.

This work was supported in part by the National Science Foundation
under Grant No.  DMR-9971348 (Z. F.).  A portion of this work was
performed at the National High Magnetic Field Laboratory, which is
supported by NSF Cooperative Agreement No.  DMR-9527035 and by the
State of Florida.  Work at Los Alamos was performed under the auspices
of the U.S. Dept. of Energy.

\begin{table}[tbp]
\caption{Structural parameters of primitive tetragonal CeRhIn$_5$.
Calculated parameters were obtained using GGA and brillouin zone convergence
as discussed in the text. The parameter $x$ is the Wykoff coordinate of
displacement of face-centered In atoms along $z$.  Experimental values 
from Hegger \textit{et al.}}
\label{band-table}
\begin{tabular}{lll}
\multicolumn{2}{l}{Space Group P4/mmm (123)} &  \\
V/V$_{exp}$ & 1.023 & (V$_{exp}$ = 163.4 \AA$^{3}$) \\ \hline
$c/a$ & 1.626 & (c/a$_{exp}$ 1.620) \\
$x$ & .302 & x$_{exp}$ .306 \\
B (V$_{exp}$) & 87 GPa &  \\
B (V$_{cal}$) & 78 GPa &  \\
Dos(E$_F$) & 10.8 eV$^{-1}$ &
\end{tabular}
\end{table}

\begin{table}[tbp]
\caption{Measured and calculated dHvA frequencies for CeRhIn$_5$ with
$\vec{B}$ along [100].  Frequencies listed for Cornelius \textit{et al.}
have been corrected by 4.5 percent (private communication).}
\label{bandfreqs100}
\begin{tabular}{ccccc}
calc. F (T) & band & F (T) & Symb. & F (T)\cite{Cornelius} \\ \hline
2139 & 92 (h) & 2176 & F$_{7}$ &  \\
849 & 92 (e) & 874 & F$_{6}$ & 861 \\
761 & 92 (e) & 722 & F$_{5}$ & 714 \\
551 & 92 (h) & 446 & F$_{4}$ & 492 \\
362 & 90 (h) &  &  &  \\
342 & 90 (h) &  &  &  \\
337 & 90 (h) & 324 & F$_{3}$ & 295 \\
283 & 94 (e) &  &  &  \\
245 & 90 (h) &  &  &  \\
214 & 94 (e) & 212 & F$_{2}$ & 219 \\
175 & 94 (e) &  &  &  \\
110 & 94 (e) & 106 & F$_{1}$ & 105 \\
26 & 92 (h) &  &  &  \\
22 & 90 (h) &  &  &  \\
21 & 92 (h) &  &  &  \\
10 & 90 (h) &  &  &  \\
9 & 90 (h) &  &  &
\end{tabular}
\end{table}

\begin{table}[tbp]
\caption{Measured and calculated dHvA frequencies for CeRhIn$_5$ with 
$\vec{B}$ along [001].}
\label{bandfreqs001}
\begin{tabular}{ccccc}
calc. F (T) & band & meas. F (T) & Symb. & F 
(T)\cite{Cornelius,CorneliusPC} \\ \hline
12326 & 92 (e) & 12280 & F'$_{10}$  &  \\
12126 & 92 (e) &  &  &  \\
11268 & 94 (h) &  &  &  \\
6878 & 92 (e) & 7200 & F'$_{9}$ &  \\
5562 & 94 (e) & 6120 & F'$_{8}$ & 6256 \\
4502 & 94 (e) & 3600 & F'$_{7}$ & 3605 \\
4331 & 94 (e) & 2740 & F'$_{6}$ & \\
3951 & 94 (e) & 2160 & F'$_{5}$ & \\
686 & 90 (h) & 1412 & F'$_{4}$ & \\
597 & 90 (h) & 562 & F'$_{3}$ &  \\
585 & 90 (h) & 216 & F'$_{2}$ & \\
451 & 90 (h) & (115) & F'$_{1}$ & %
\end{tabular}
\end{table}

\begin{table}[tbp]
\caption{Measured masses for dHvA frequencies for CeRhIn$_5$ with $B$ along [100].
Our data also are compared to that of Cornelius \textit{et al.}}
\label{cmass100}
\begin{tabular}{ccccc}
Symb. & F (T) & m* ($m_{e}$) & F (T)\cite{Cornelius} &  m* ($m_{e}$) 
Ref. \cite{Cornelius} \\ \hline
F$_{6}$ & 874 & 3.3 $\pm$ 0.6 & 861 & 1.31 $\pm$ 0.22 \\
F$_{5}$ & 722 & 2.0 $\pm$ 0.8 & 714 & 1.17 $\pm$ 0.09 \\
F$_{4}$ & 446 & 3.9 $\pm$ 0.9 & 492 & 0.72 $\pm$ 0.14 \\
F$_{3}$ & 324 & 5.5 $\pm$ 1.0 & 295 & 0.93 $\pm$ 0.15 \\
F$_{2}$ & 212 & 2.9 $\pm$ 0.4 & 219 & 0.99 $\pm$ 0.10 \\
F$_{1}$ & 106 & 3.3 $\pm$ 0.6 & 105 & 1.36 $\pm$ 0.23
\end{tabular}
\end{table}

\begin{table}[tbp]
\caption{Measured dHvA frequencies for CeRhIn$_5$ with $B$ along
[001].  Frequencies marked with asterisks appear only at the lowest
temperatures.  Calculation of masses for the frequencies marked with
daggers was not possible because the amplitudes did not change
significantly over the measured temperature range.  The frequency in
parentheses is weak due to a low number of oscillations in the sweep
range.}
\label{cmass001}
\begin{tabular}{ccccc}
Symb. & F (T) & m* ($m_{e}$) & F (T)\cite{Cornelius} &  m* ($m_{e}$) 
Ref. \cite{Cornelius} \\ \hline
F'$_{10}$ & 12280* & -- & & \\
F'$_{9}$ & 7200* & -- &  &  \\
F'$_{8}$ & 6120 & 6.1 $\pm$ 0.3 & 6256 & 6.5 $\pm$ 0.8 \\
F'$_{7}$ & 3600 & 4.6 $\pm$ 1.0 & 3605 & 4.8 $\pm$ 0.4 \\
F'$_{6}$ & 2740* & -- & & \\
F'$_{5}$ & 2160* & -- & & \\
F'$_{4}$ & 1412 & 4.5 $\pm$ 0.3 & & \\
F'$_{3}$ & 562\dag & -- &  & \\
F'$_{2}$ & 216\dag & -- & & \\
F'$_{1}$ & (115)\dag & -- &  &
\end{tabular}
\end{table}

\end{document}